# Investigating 3D Printer Residual Data

Daniel Bradford Miller
Computer Science
School of Computing
University of South Alabama
dmiller@southalabama.edu

Jacob Gatlin
Computer Science
School of Computing
University of South Alabama
jrg1222@jagmail.southalabama.edu

William Bradley Glisson
Cyber Forensics Intelligence Center
Department of Computer Science
Sam Houston State University
glisson@shsu.edu

Mark Yampolskiy
Computer Science
School of Computing
University of South Alabama
yampolskiy@southalabama.edu

Jeffery Todd McDonald
Computer Science
School of Computing
University of South Alabama
jtmcdonald@southalabama.edu

## Abstract

*The continued adoption of Additive Manufacturing (AM) technologies is raising concerns in the security, forensics, and intelligence gathering communities. These concerns range from identifying and mitigating compromised devices, to theft of intellectual property, to sabotage, to the production of prohibited objects. Previous research has provided insight into the retrieval of configuration information maintained on the devices, but this work shows that the devices can additionally maintain information about the print process. Comparisons between before and after images taken from an AM device reveal details about the device's activities, including printed designs, menu interactions, and the print history. Patterns in the storage of that information also may be useful for reducing the amount of data that needs to be examined during an investigation. These results provide a foundation for future investigations regarding the tools and processes suitable for examining these devices.*

## 1. Introduction

Additive Manufacturing (AM), also known as 3D printing, is a manufacturing technique that constructs objects by adding consecutive layers of material. Gartner [1] predicts that the increasing usage of AM technologies will enable the creation of new business models. In addition, the report predicts increased penetration into consumer manufacturing, aerospace, and healthcare industries. From a local market perspective, consumer 3D printers allow home users to produce a variety of objects based on designs downloaded via Internet repositories, purchased from third-party designers, or created by the user using Computer Aided Design (CAD) software. A 2017 report by Arizton Advisory & Intelligence's [2] forecast that the 3D Printing Market revenues will surpass $11 billion by 2022.

While this technology has significant potential to enhance both the capabilities and reach of manufacturing, the potential for misuse is a cause for concern. Fruehauf et al. [3] note that the flexibility of this family of manufacturing processes permits the creation of a wide variety of objects, from trinkets and custom components to Automated Teller Machine (ATM) skimmers, weapons, and illicit chemicals. Several researchers have also voiced apprehensions about the technology enabling infringement on Intellectual Property (IP) [4-6]. In addition, security researchers investigating these technologies have successfully demonstrated means of subverting the devices [7, 8], the data files and communications protocols involved in the manufacturing process [9, 10], and have even considered the use of the AM system itself as a weapon [11]. The potential for misuse coupled with the vulnerabilities present in the underlying devices contributes to the likelihood that these devices will need to be examined in a forensic manner to assist corporate, civil, and criminal investigations.

A growing concern in the digital forensics community is the volume and variety of data collected during an investigation. Garfinkle [12] identified the increasing size of storage media and increasing connectedness of systems as challenges for forensic practitioners. Shaw and Atkins [13] also found that forensic analysis of embedded devices was often dependent on data residing on other systems connected to the operation of the device. Miller et al. [14] found a theme of connectedness in the workflows for AM

HICSS

systems, particularly in systems which offered greater functionality. Quick and Choo [15] surveyed the research addressing these challenges and found significant deficits in the use of data reduction techniques, data mining, and intelligent analysis. Tassone et al. [16] endorsed the use of visualizations to attempt to reduce both the data volume and cognitive load. These findings coupled with the ever-increasing evolution of technology capabilities forewarn that investigations including AM systems are likely to be complex and time-consuming, necessitating the development of data reduction strategies.

This atmosphere prompts the hypothesis that AM devices can be profiled after legitimate user interactions from a residual data perspective. To address this hypothesis, the following research questions were identified: Can data be extracted from an AM device? If so, does that data contain residual data of the print operations? Are current digital forensics tools able to retrieve and process this data? Can the volume be reduced in a forensically sound manner?

The contribution of this research is two-fold. First, it provides an empirical demonstration of the viability of residual data on a specific 3D printer. In doing so, it documents artifacts that could be useful in an investigation. Second, it provides the foundation for future investigations regarding the tools and processes suitable for examining these devices. This paper structure is as follows: Section 2 presents related work, Section 3 presents the method for the exploratory examination and experiment. Section 4 presents results and analysis, Section 5 discusses the results, and Section 6 presents the conclusions and details future work.

## 2. Related Work

Widespread use of embedded systems prompted research into systems [13] [17], and the residual data they contain [18]. The continued incorporation of AM technologies into commercial manufacturing processes and the rapid development of manufacturing in the private sector indicates AM will follow a similar trend of increasing ubiquity. This trend is encouraging researchers to look into 3D printers to proactively develop methods for examining AM devices and systems in a forensic context [14] [19].

According to Shaw and Atkins [13], an embedded system is a non-user-programmable computer performing a few dedicated functions. They highlight access to the data storage media as a major challenge to forensic analysis in these systems. They state that many embedded systems include data storage components as integral parts of the device, which can make the removal of the data storage for traditional forensic image acquisition difficult and increase the risk of altering or destroying data.

Despite the difficulties presented by these devices, researchers have detailed the investigations of embedded systems. Van Vliet et al. [17] described the investigation of a ground-level controller after a fire in a wind turbine. They noted several obstacles to acquiring the log entries from the device, including the need to supplement the onboard batteries to maintain the logs and the decision to recreate a portion of the controller's logical environment prior to log retrieval. Log retrieval involved using a manufacturer provided tool to access the stored data. The process and results of the Vliet's case study are representative of the investigation techniques examined by Shaw and Atkins [13]. Both have a component of known origin and involved significant collaboration with the manufacturer of the device.

Grispos et al. [18] examined smartphones as a proxy for forensic analysis of cloud storage services. The authors' experimental methods centered on preparing smartphones with cloud storage applications linked to accounts with experimental files, then performing well-defined file manipulations. Residual data from these manipulations was captured from both internal memory and an SD memory card. The results showed a large amount of residual and metadata was left intact when files were deleted from the local device, and that many manipulations had identifiable residual effects. The authors conclude that smartphones can serve as a forensic proxy for cloud storage services, even with a black-box view of the application in question.

Unfortunately, not all embedded systems originate from known a device manufacturer. Souvignet et al. [20] detail the investigation of an ATM card skimmer located by police in the European Union. The device in question was constructed from several commodity hardware components. The authors documented a combination strategy for their analysis. The first stage of their investigation consisted of a black box analysis in which they examined the hardware to determine likely device capabilities. From the results of that analysis, the decision was made to physically deprotect the microcontroller to facilitate access to the device firmware. From the firmware, they were able to reverse engineer the encryption routine and recover most of the information the device contained. They go on to describe an Android application created to enable the detection of the devices via Bluetooth radio. This work shows not only a successful device analysis but also that the applicability of forensic analysis is not limited to courtroom environments.

Miller et al. [14] classified 3D printers in terms of functionality and the requirements for external control. Utilizing open source intelligence garnered from the manufacturer-produced documentation, they identified patterns in the processes used by their classifications to move data before and during print operations. They found evidence in the description of those processes that devices capable of independent operation and devices that offering the capability for local design storage were very likely to have residual data of the print process on the device. The authors called for further research to verify their findings.

Garcia and Varol [19] examined the internal hard disk drive of an Object24, a 3D printer manufactured by Stratasys. They were able to image the disk through a commodity USB adapter and analyzed the image with Guidance Software's EnCase Forensic Suite. They were able to recover network configuration data, logs, and device settings from the image, but did not locate design files or pictures of the printed objects in the acquired image. This work represents an initial attempt at analyzing an AM device. Further examination of the image using additional tools and search strategies would enhance the results.

Current research demonstrates that other types of embedded systems have undergone forensic analysis, but researchers are just starting to investigate devices that associated with AM technologies as potential sources of residual data. Currently, there is minimal research that focuses on acquiring and analyzing AM devices from a forensics perspective.

## 3. Methodology

This work takes the form of an exploratory case study as described by Oates [21]. To investigate the hypotheses and associated research questions, a two-stage approach was adopted based on the strategy developed by Souvignet and Frinken [22]. The first stage consists of a black box analysis with the goals of identifying hardware capabilities, data storage locations, and potential access methods. The results of the first stage influenced the approach used for the second stage. A grey box investigation strategy was selected due to the availability of an open source distribution [23] of a documented underlying operating system [24] along with documentation on the implementation of an open source application stack [25]; however, nominal information regarding manufacturer customizations to that distribution. These conditions fulfilled the criteria of incomplete system knowledge as utilized by Jehan, Pill, and Wotawa [26] to qualify a study as 'grey box'.

A single Voxel8 3D printer was selected for this investigation. Specific printer information is provided in Table 1. This printer is representative of a class of printers identified by Miller, et al. [14] as likely to contain residual data from the print process. Features qualifying this printer for the classification include the ability to print objects without a controlling machine (standalone operation) and the presence of internal storage for g-code files. Two software packages were used for analysis in this research: AccessData's Forensic Tool Kit (FTK) version 6.2 [27] and Autopsy version 4.1.1 [28]. FTK is a commercial forensics platform frequently accepted by U.S. Courts as a forensically sound tool. Autopsy is an open source forensics platform. These platforms were selected due to their ability to filter results based on a set of file hashes and license availability at the time of the study.

**Table 1: Voxel8 Specifications**

| Printing Technology | Fused filament fabrication (FFF), Pneumatic Direct Write |
|---|---|
| Build Volume | 150 mm x 150 mm x 100 mm |
| Layer Resolution | 100 microns |
| Conductive Trace Width | 250 microns |
| Filament Size | 1.75mm |
| Materials | Polylactic acid (PLA), Conductive Silver Ink |
| Network Connectivity | Ethernet, Wi-Fi |

### 3.1 Black Box Examination

The equipment used for this work consisted of three computers and two media interfacing components. A workstation was selected to control the printer during the experiment, further referred to as the User PC. Interactions with the printer were conducted via the Google Chrome web browser. A second workstation was selected to perform image acquisitions and analysis, further referred to as the Forensics PC. A third machine was used to restore image files to the SD card. An Insignia USB SD Card Reader (Model NS-DCR30S2K) was used to interface the SD card to the forensics PC and the imaging PC for both read and write operations. Read operations involving the SD card were conducted via a Wiebtech USB write blocker (Model 31300-0192-0000) to prevent modifications to the data by the forensics PC. Configuration details for the computers utilized in the research are presented in Table 2. Hardware was selected due to its availability at the time of the experiments and exceeding the system requirements for Google Chrome (User PC), Autopsy and FTK (Forensics PC), and *dd* (Imaging PC).

A physical inspection of the Voxel8 was conducted to locate and examine the electronic components of the machine. The machine was turned off during the examination. Two commodity components were identified behind the rear access panel that constituted much of the circuitry in the machine. These components consisted of a Raspberry Pi 2 Model B v1.1 and a RepRap Arduino-compatible MotherBoard (RAMBo) v1.3.

**Table 2: Hardware**

| User PC | OS: Windows 7 SP1 64-bit<br>Processor: Intel Xeon e5-1607v4<br>RAM: 16GB |
|---|---|
| Forensics PC | OS: Windows 7 SP1 64-bit<br>Processor: Intel Xeon e5-1650v4<br>RAM: 32GB |
| Imaging PC | OS: Ubuntu 16.04.3 LTS<br>Processor: Intel Core i7-3770<br>RAM: 8GB |
| SD Card Reader | Insignia Model NS-DCR30S2K |
| Write Blocker | Wiebtech Model 31300-0192-0000 |

It was determined that the Raspberry Pi provided the control interface for the machine. This deduction was due to a High-Definition Multimedia Interface (HDMI) cable and Universal Serial Bus (USB) cable connected to the Raspberry Pi and routed to the front panel. Additionally, the external network port on the Voxel8 was connected to the Raspberry Pi. Due to the cabling connecting the RAMBo to the motors controlling the gantry and extruder assembly, it was determined that the RAMBo provided control signals responsible for the operation of the printer.

The Raspberry Pi and the RAMBo were connected via a USB cable. The Raspberry Pi had a Micro Secure Digital (SD) card inserted into its card reader. No other storage devices were noted within the chassis. The rear compartment of the device is pictured in Figure 1. The Raspberry Pi is the green printed circuit board (PCB) on the left, and the RAMBo is the green PCB on the right in the image.

The micro SD card was imaged for further analysis. The following process was adopted for image acquisitions to ensure the data was not corrupted and the acquisition process did not alter the data on the SD card:

1. Power down the Voxel8
2. Turn off the power switch
3. Remove the Micro SD card from the Raspberry Pi and insert it into a Micro SD to SD adapter

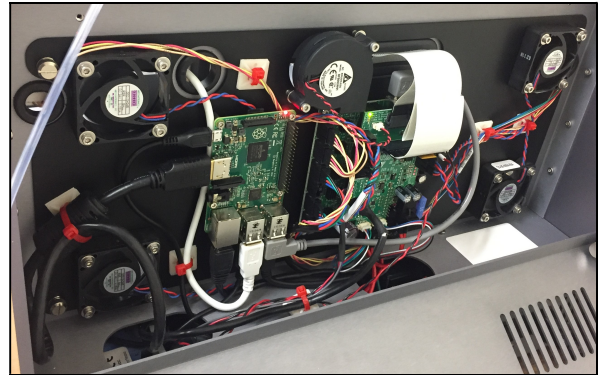

**Figure 1: Rear Compartment of the Voxel8**

4. Connect the Micro SD card adapter to the forensics PC via the USB write blocker.
5. Create a RAW disk image from the MicroSD card using FTK Imager's physical acquisition option and confirm that the application reports a successful verification
6. Remove the volumes associated with the micro SD card through the operating system and disconnect the USB write blocker

FTK Imager produces both Message Digest 5 (MD5) and Secure Hash Algorithm 1 (SHA1) hashes of the media during acquisition. It then uses those hashes to verify the integrity of the acquired image. A text file containing details of the imaging process, such as time of the acquisition and the image hashes generated by FTK, was saved with the image file. The images were saved in a RAW format to facilitate restoration to the original media.

The initial image was successfully imported into Autopsy with all ingest modules selected. Exploration of the content revealed a Master Boot Record (MBR) partition scheme with 4MB of unallocated space at the beginning of the media and two defined partitions, one File Allocation Table (FAT) partition of 60MB and one Linux partition of 15,143MB. Figure 2 is a screenshot of the partition table, top-level directory trees, and hexadecimal presentation of the beginning of the initial state image shown in Autopsy.

Except for the MBR, the unallocated space at the beginning of the media contained no data (all bytes recorded as 0x00). There was no boot code present in the first 446 bytes of the MBR.

The FAT partition was formatted FAT16 and contained bootstrap code for an Advanced Reduced Instruction Set Computing (RISC) Machine (ARM) version of Linux. Also present were several Device Tree Blob (DTB) files describing a Broadcom hardware set and configuration files for Octopi, an ARM Debian Linux distribution built for operating 3D printers.

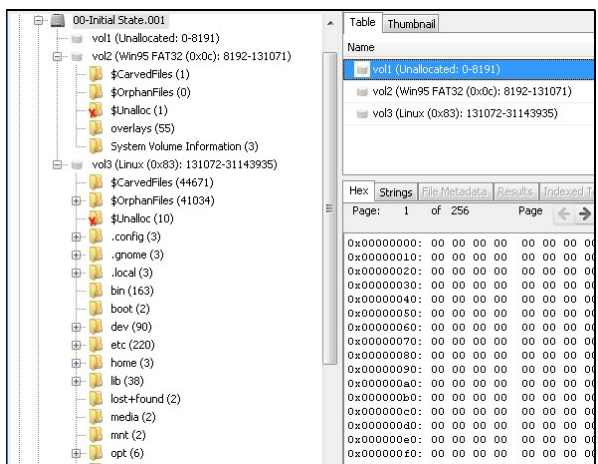

**Figure 2: SD card partitioning and file system**

The Linux partition was determined to contain an Extended File System (ext) based on the superblock location and signature. The folder structure in the root of the filesystem included folders specified by the Unix Filesystem Hierarchy Standard [29]. Files in the folder */etc* identified the operating system distribution (Octopi v0.13.0) and further identified the Debian distribution from which it was derived (Raspbian Jesse release 2016-02-09).

### 3.2. Grey Box Testing

A series of one group, pre-test, post-test experiments as described by Oates [21] was developed to assess changes in the state of the files stored on the Voxel8's SD Card as a result of user actions. Comparison of file content was conducted by comparing Message Digest 5 (MD5) hashes calculated from the individual files. Differences in MD5 file hashes indicate a difference in the hashed content. The dataset used for this experiment consisted of a single new G-code file titled Forensic_Test_Print.gcode and a single G-code file selected from files already present on the Voxel8 titled Rectangular_Test_Print.gcode. MD5 hashes for the G-code files were generated and used to create an alert hash set for use for FTK's KFF. A second hash set was generated from the MD5 hash values of all the files present in the image from the micro SD card extracted from the Voxel8. The second hash set was used to establish an ignore hash set for the KFF.

The experiment was divided into an eleven-stage process. The process was repeated for each state manipulation sequence examined.
1. Restore the initial image acquired from the micro SD card during the black box examination to the micro SD card and verify the restored media's integrity by comparing the block device's MD5 hash to that recorded by FTK Imager for the image.
2. Insert the micro SD into the Raspberry Pi
3. Power on the Voxel8 printer
4. Perform state manipulations
5. Power down the Voxel8 using the devices front panel display.
6. Turn off the power switch
7. Remove the Micro SD card from the Raspberry Pi and insert it into a Micro SD to SD adapter
8. Connect the Micro SD card adapter to the forensics PC via the USB write blocker
9. Create a RAW disk image from the MicroSD card with FTK Imager's physical acquisition option and confirm that the application reports a successful verification.
10. Remove the volumes associated with the micro SD card through the operating system and disconnect the USB write blocker
11. Import the image into FTK

### 3.3. State Manipulations

The process resulted in an image documenting the state of the device after each state manipulation sequence. The manipulation sequences were as follows:
1. Upload a File: Log into the Voxel8's OctoPrint web server as 'admin' and upload a design titled 'Forensics Test Print'.
2. Upload and Print a File: Log into the Voxel8's OctoPrint web server as 'admin' and upload a design titled 'Forensics Test Print'. Start the print by selecting the 'Start Print' prompt on the page and accepting the default options. Allow the job to complete and respond 'Yes' to the 'Print completed successfully' option on the front panel.
3. Delete an Existing File: Log into the Voxel8's OctoPrint web server as 'admin' and delete the file 'Rectangular Test Token' from the Voxel8
4. Upload and Delete a File: Log into the Voxel8's OctoPrint web server as 'admin' and upload a design titled 'Forensics Test Print'. Then delete the file 'Forensics Test Print' from the Voxel8.
5. Upload, Print, and Delete a File: Log into the Voxel8's OctoPrint web server as 'admin' and uploaded a design titled 'Forensics Test Print'. Start the print by selecting the 'Start Print' prompt on the page and accepting the default options. Allow the job to complete and responded 'Yes' to the 'Print completed successfully' prompt on the front panel. Delete the file 'Forensics Test Print' from the device through the OctoPrint web server.

Page 7180

6. Cancel a Print: Log into the Voxel8's OctoPrint web server as 'admin' and start a print for the file 'Rectangular Test Token' by selecting 'Start Print' in the web interface and accepting all the default options. When printing starts, press the 'X' button, on the front panel to cancel the print job. When prompted, select 'Print Problem' on the front panel as the reason for halting the job.
7. Cancel and Delete a Print: Log into the Voxel8's OctoPrint web server as 'admin' and uploaded a design titled 'Forensics Test Print'. Start the print by selecting 'Start Print' on the page and accepting the default options. When printing starts, press the 'X' button on the front panel to cancel the print job. When prompted, select 'Print Problem' on the front panel as the reason for halting the job. Then delete the file 'Forensics Test Print' from the device through the OctoPrint web server.
8. Printing and Canceling using the Front Panel: Initiate a print of the design 'Rectangular Test Token' from the front panel. When printing starts, press the 'X' button, on the front panel, to cancel the print job. When prompted, select 'Print Problem' on the front panel as the reason for halting the job.
9. Printing, Canceling and Deleting using the Front Panel: Initiate a print of the design 'Rectangular Test Token' from the front panel. When printing starts, press the 'X' button on the front panel to cancel the print job. When prompted, select 'Print Problem' on the front panel as the reason for halting the job. Then delete the file 'Rectangular Test Token' from the device through the front panel menu.
10. Update Printer Firmware through OctoPrint: Instruct the Voxel8 to download and install a firmware update through OctoPrint's web interface. The front panel displays an 'Update Complete' upon task completion.

### 3.4 Data Processing

Per the Computer History Model proposed by Carrier and Spafford [30], the differences between the images are the result of events which occurred on the device between the image acquisitions. Cryptographic hash values were utilized to determine whether individual files were changed between the initial and post-manipulation images. The hash sets were loaded as custom hash sets into the KFF server component of FTK. The hash set containing the known G-code files was configured as alert file set, and the hash set containing all files extracted from the initially acquired image was configured as a known file set. The results presented by FTK after hiding the known files were filtered to include only allocated, non-deleted files. These files were flagged for manual file inspection to determine their content and included both changed and newly created files with unique content. A specific search was conducted for the G-code file 'Rectangular Test Token' by its hash value against the initially acquired image and the file locations were recorded. These file locations were searched for in the images to determine the state of those files due to overlap in the hash sets.

### 3.5. Limitations

This study examines a single instance the Voxel8 printer, but seventy-one (71) consumer 3D printer models are listed on OctoPrint's supported printers page and six of those models utilize the hardware combination of a Raspberry Pi and a RAMBo [31]. It should be noted that there were no visible or documented means to reset the machine to a factory default setting. Testing of this machine required connectivity to the general Internet and there is the possibility that the machine was subjected to network traffic not included as part of the experiment. Data carving and event-correlation are considered out of scope for the purposes of this work.

### 4. Results and Analysis

Differences in the file systems were detected each experiment. Table 3 summarizes the number of differences identified between the initial and post manipulation images. Differences include file changes, creations, and deletions. Ninety (90) unique files were altered by ten (10) experiments. Uniqueness was defined by file location and base file name.

Table 3: File system changes by experiment

| Manipulation set | Changes |
|---|---|
| Upload a File | 52 |
| Upload and Print a File | 66 |
| Delete an Existing File | 51 |
| Upload and Delete a File | 59 |
| Upload, Print, and Delete a File | 61 |
| Cancel a Print | 50 |
| Cancel and Delete a Print | 49 |
| Printing and Canceling using the Front Panel | 50 |
| Printing, Canceling and Deleting using the Front Panel | 51 |
| Updating Printer Firmware through OctoPrint | 80 |
| Unique files changed | 90 |

## 4.1. Manual File Inspection

Manual inspection of the files identified by FTK revealed several instances of data related to the print process. A summary of notable changes follows:

*/home/pi/.octoprint/logs/octoprint.log*: The log retained interactions with the OctoPrint server such as the Internet Protocol (IP) addresses of clients connecting to it as well as system level commands initiated by the user. For manipulations including the upload of a file, a reference to the name of the design with an appended numeric identifier and timestamp were found in the log. In some cases, a log rotation mechanism split the log file. There were not any unique identifiers noted for users or external devices.

*/home/pi/.octoprint/uploads/.metadata.yaml*: The file containing a list of uploaded G-code files. The device recorded attributes of those files, including file name, the SHA1 hash of the file, estimated print time and filament usage, a numeric identifier for the print file, and the time and status of the last print attempt. In runs that included a delete manipulation, the data for the deleted file was not present in the list.

*/home/pi/.octoprint/uploads/Forensics_Test_Print-224546099-<UTC Timestamp>.gcode*: A copy of the 'Forensic Test Print' file uploaded to the device was found in this location. The integrity of the file was verified by SHA1 and MD5 hash values that are identical to the hash values of the file uploaded to the Voxel8 during the experiments. The file was only present during state manipulations that included file upload and did not include a delete operation.

*/home/pi/.octoprint/uploads/Rectangular_Test_Token-770373878-2017-09-11T21-27-35.303Z.gcode*: A copy of the design 'Rectangular Test Token' file was identified at this location. The file was not present in images taken after manipulation sequences that included a step to delete that design file. The hash value for this file was also associated with a file in the *ext* partition named */root/.cache/chromium/Default/Cache/F_000016*.

*/root/.cache/chromium/Default/Cache/data_3*: A binary file containing browser cache data from Chromium was identified. The file contained a copy of the uploaded G-code file intermingled with additional information. This data remained on the machine during experimental runs that included deletion of the uploaded file.

*/root/.cache/chromium/Default/Cache/F_000017*: A copy of the uploaded G-code file "Forensic Test Print" was retrieved from this location. The content of the file was verified against the file uploaded to the Voxel8 by SHA1 and MD5 hash values. The file was present after all upload operations and remained on the machine after the design was deleted.

*/root/.config/chromium/Default/CurrentSession*: A binary file that appeared to contain references to the states of the printer shown on the front panel interface of the device was retrieved from this location. References were extracted from the file by excluding non-printable characters.

*/root/.config/chromium/Default/History*: An SQLLite database file containing URLs that mirrored the navigation of the device conducted via the front panel interface was retrieved from this location. The entries contained labels such as 'menu:print' and 'menu:delete' with timestamps present in cases where the item was selected.

## 4.2 Aggregate Change Analysis

Examining the distribution of changes reveals details about the operation of the device. Figure 3 shows a heat map of the number of changes by directory and normalized by directory. Directories which did not have a change occur are excluded from the heat map. No changes were identified in the FAT partition. Changes that occurred within the *ext* partition were constrained to five top-level directories: */etc*, */home*, */root*, */tmp*, and */var*. The heat map shows variation in the numbers of changed files in */home/pi/.octoprint*, */root/.cache/chromium/Default/Cache*, and */root/.config/chromium*. The map also indicates differences where the changes occurred depending on the manipulation. For instance, the firmware update affected files in different directories than non-administrative operations.

## 5. Discussion

The changed files identified in this work included metadata concerning the print operations, the G-code files, and device log files. The presence of this information has implications for the forensic, security, and privacy issues surrounding the device.

On the forensic front, this device serves as a potential source of information during an investigation. The residual data contains several references to what occurred on the device, namely what was printed, when it printed, and the state of that print job. While legitimate interactions did result in the removal of the metadata for the print files, copies of the G-code appeared in several file system locations, and those copies maintained in the Chromium cache persisted after deletion. In the absence of the means to reset the device, this information would persist until the cache was cleared due to excessive size. There were no instances of cache clearing observed during this experiment.

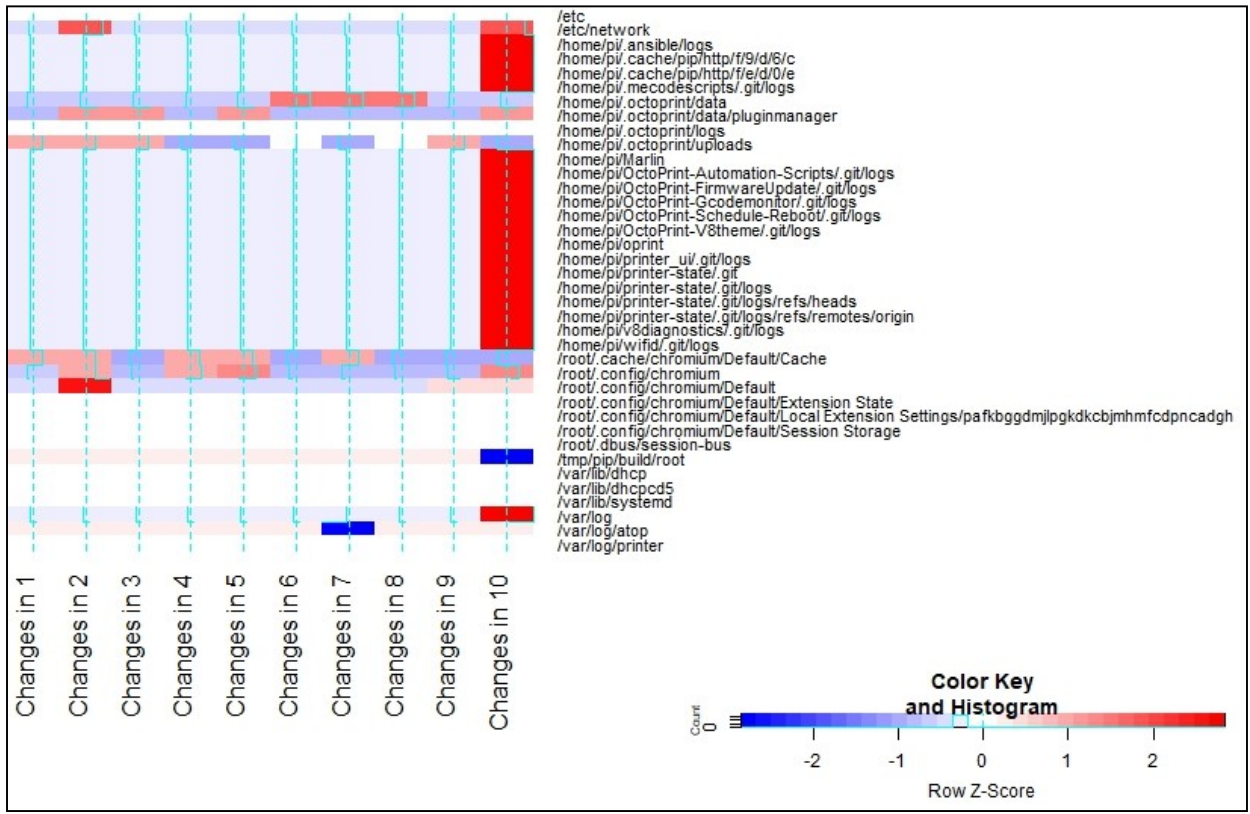

**Figure 3: Heatmap of changes by directory and experiment run**

It was also noted that the device did not maintain explicit entries recording the deletion of print files from the machine. In this experiment, file deletion was inferred from the difference between the acquired images. Because the files are maintained in the Chromium cache, purposeful deletion can be inferred by correlating the absence of the file in /home/pi/.octoprint/uploads with files existing in /root/ .cache / chromium / Default/Cache.

The chassis used for the device was not secured and commodity hardware was used in its construction. There was also no encryption used on any of the altered files, nor was there encryption applied to either the *ext* or FAT volumes. From a security perspective, the ease of access makes the device very difficult to secure. Consequently, protecting the data contained on the device from malicious actors it is difficult. From a forensic standpoint, these factors reduced the complexity of data acquisition and analysis. All that was necessary to access the data in this case was to remove the SD card and use the acquisition and analysis tools as they were intended.

This work provides contrast to other examinations of AM devices that exist in the literature. In both this work and Garcia and Varol [19], tools commonly accepted in U.S courts were used, but the differential techniques used in this examination were able to locate residual data of interest, specifically the g-code files for the printed models, not located in previous research, even in cases where the data of interest were intermingled with other data. Similarly, the differential techniques did not identify files with static configuration data, such as the network configuration or authentication keys. Hybrid techniques are likely to provide the best picture of a device's history.

The change patterns illustrated in the heat map as illustrated in Figure 3 indicate the files with which the software interacted varied depended on the activities which occurred. The data collected was of insufficient size to perform statistical modeling of the interactions between the actions and observed changes. The visually inferred variance from the heat map prompted additional investigations into which files were responsible. In the case of */root/.config/chromium*, a substantial number of changes were due to Chromium updating whitelists used for its Safe Browsing feature. These updates did not occur during every test, and there was no activity prescribed by the test protocol intended to cause such an update. For */home/pi/.octoprint/uploads*, the variance resulted from the addition of the g-code files and modification of *metadata.yaml*, which resulted from actions prescribed by the experimental method. Consequently, a usage profile based on aggregate changes would require a manual review to identify changes captured that were not relevant to the activity of interest. In the context of this device, the most relevant directories were found to be */home/pi/.octoprint/uploads*,*/home/pi/.octoprint/logs* and */root/.cache/chromium/Default/Cache*.

## 6. Conclusions and Future Work

This work demonstrates a successful retrieval of data from a specific AM device. The device utilized commodity flash storage to contain the operating system for the device providing interface and control functionality. This hardware configuration permitted off-the-shelf equipment to be used to acquire data from the device. The use of a common partition table and file system did not require any special software to interpret the image. For this device, there were no complications introduced into the acquisition processes by how the data was stored

This work also found that files changed during operation of the Voxel8's contained residual data concerning the print process. The residual data included G-code files and metadata detailing the time actions were performed on the device. There were notable holes in the metadata, specifically related to authentication. This data would be suitable for inferring that something occurred and providing some information useful for identifying where the event was initiated. However, little in the changed data could be positively linked to static configuration parameters, such as keys, certificates, or user identifiers.

The tools selected for this work were capable of ingesting and interpreting the images acquired from the Voxel8. The base functionality of the software, that of storage media image ingest and the ability to filter results based on a hash database, were sufficient to conduct the analysis detailed in this work. The use of the KFF did reduce the number of files subjected to manual inspection, but the manual file inspection process was still long and arduous.

Changes were only detected in five out of the 21 top-level directories of one volume during this experiment. Of the active top-level directories, a similar pattern of changes was only present in a small subset of directories that persisted. This result could be of significant use to forensic investigators in device triage situations, and security practitioners for identifying anomalous activity. Both activities would require an authoritative profile source to guide their actions. This source could be either the manufacturer of the device or a trusted organization.

This work shows that data can be extracted from an AM device. It also shows that the data extracted from the device did contain information about the operation of the printer. Further, this work used open source and commercial forensics platforms to ingest and process acquisitions from the device. Using the filtering capabilities of these platforms, a method was designed which resulted a reduction in the number of manual file inspections to a handful of files without full knowledge of how the system operates. Based on these results, it is concluded that it is possible to profile an AM device to locate residual data relevant to legitimate user activities.

Future work will look at performing similar analysis on other AM platforms. These efforts will identify emerging trends in the system architectures present in these devices and identify common acquisition and analysis techniques for practitioners. Additionally, further development of profiling methods will be conducted to explore the use of statistical modeling techniques to correlate activities with expected changes to the file system. Further work is also necessary to examine the correlation between events recorded by the device with events that occur on other devices in AM systems.